\documentstyle[prl,aps,eqsecnum]{revtex}
\begin{document}
\draft
\input epsf

\newtheorem{defn}{Definition}
\newtheorem{intro}{Introduction}

\newcommand{\bs}{$\backslash$}
\newcommand{\comname}[1]{{\sf \bs #1}}
\newcommand{\envname}[1]{{\sf #1}}
\newcommand{\be}{\begin{equation}}
\newcommand{\ee}{\end{equation}}
\newcommand{\ben}{\begin{eqnarray}}
\newcommand{\een}{\end{eqnarray}}
\newcommand{\oz}{{\cal{A}}}
\newcommand{\la}{\lambda}
\newcommand{\th}{\theta}
\newcommand{\no}{\noindent}

\title{Non-commutative Geometry and the Higgs Masses}

\author{ Aybike \c{C}atal\footnote{E.mail: catal@metu.edu.tr}
     and
    Tekin Dereli\footnote{E.mail: tdereli@metu.edu.tr} }
\address
{Department of Physics,
  Middle East Technical University , 06531 Ankara, Turkey}

\maketitle

\bigskip

\bigskip

\begin{abstract}
We study a non-commutative generalization of the standard electroweak
model
proposed by Balakrishna, G\"{u}rsey and Wali
[Phys. Lett. {\bf B254}(1991)430]
that is formulated in terms of the derivations $Der_{2}(M_3)$ of a
three-dimensional representation of the $su(2)$ Lie algebra of weak
isospin.
The linearized Higgs field equations and the scalar boson mass
eigenvalues are explicitly given.
A light Higgs boson with mass around 130 GeV together with four very
heavy
scalar bosons are predicted.

\end{abstract}
                                \bigskip

\section{Introduction}

In spite of its observational successes, the standard model of
electroweak
interactions cannot yet be considered as a fundamental theory
because the scalar boson sector, unlike the gauge sector involving
the fermions and the gauge bosons, has to be
written down in an ad hoc way and not by gauge principles.
Furthermore, the unavoidable Higgs scalar has not
been observed and there is no way to predict its mass.
In this connection a remarkable attempt at unifying
gauge fields and Higgs scalars was suggested by A. Connes
\cite{connes1},
making use of the tools of non-commutative differential geometry.
The formalism involves three steps: First, a spectral triplet
$(\cal{D}, \cal{H}, \cal{A})$ is introduced, consisting of
the (generalized) Dirac operator $\cal{D}$ that acts on a
Hilbert space of states $\cal{H}$, together with an associative
$C^*$-algebra $\cal{A}$ also acting on $\cal{H}$.
Next, $\cal{A}$ is related with the algebra of complex valued functions
on space-time in the commutative case, whereas in more complicated
settings
in which the gauge groups are non-Abelian, $\cal{A}$ has to be
replaced by the tensor product
${\cal{A}} = C^{\infty}(V) \otimes M_{n} $
with an appropriate  matrix algebra. Finally, the construction of
Yang-Mills Lagrangian is done by replacing the Dixmier trace
instead of integration.
Within the above scheme,
a generalization of the standard electroweak model in
non-commutative geometry can be given
as a gauge theory with a built in spontaneous symmetry
breakdown mechanism. This way, it is not only
the Higgs sector that arises naturally, but also the correct hypercharge

assignments acquire a natural meaning.
The earliest model  along these lines is due to Connes and Lott
\cite{connes2}.
Several other attempts followed since then
\cite{violette1} , \cite{coquereaux}, \cite{chamseddine}.
Here we wish to re-examine the Higgs masses in a model proposed by
Balakrishna, G\"{u}rsey and Wali (BGW) \cite{feza2}.
In this approach the Yang-Mills and Higgs fields
occur on equal footing and the Higgs potential
consisting of a sum of complete squares, appears already
shifted onto an absolute minimum.
Thus, both the gauge boson and Higgs boson masses
can be predicted in terms of two mass scales, each related with one of
the
$SU(2)_{I} \times U(1)_{Y}$ gauge symmetry groups.

\section{Mathematical Framework}

In order to study the bosonic sector alone it is enough to deal with the

tensor product space
${\cal{A}} = C^{\infty}(V) \otimes M_{n} $ so that $\cal{A}$
can be regarded as the set of matrix valued functions on the
space-time manifold $V$ and is itself a $C^*$-algebra.
The differential calculus of this space has been studied in
\cite{madore}.
It is also possible to identify the vector fields of ${\cal{A}}$ with a
restricted set of derivations of $M_{n}$ rather than the algebra of all
derivations of $M_{n}$.
 We have this extra freedom because $Der(M_{n})$ is not a module over
$M_{n}$.
Here the Lie subalgebra $Der_{2}(M_{3})$ generated by a three
dimensional
representation of su(2) is used rather than the Lie algebra $Der(M_{3})$
of
all derivations of $M_{3}$.
Exterior derivation, connection, curvature are defined as in
\cite{feza2}, but with some modifications \cite{madore}.
The dimension of $Der_{2}(M_{3})$ is $2^{2} - 1 = 3$. Hence we may take
as the generators of $M_{3}$, the first three Gell-Mann matrices
$\tau_{1},  \tau_{2}, \tau_{3}$ and the generators of the
U-spin and V-spin subalgebras along
with the identity $\tau_{0}$ which we identify with Y $+ \frac{2}{3}$
where Y is
the hypercharge $\tau_{8} / \sqrt{3}$. The generators of the U-spin and
V-spin subalgebras are
\begin{eqnarray}
U_{\pm} & = & \frac{1}{2} \ (\tau_{6} \pm i \tau_{7}) \ , \ \
\ \ \ U_{3}  =  \frac{1}{2} \ [U_{+}, U_{-}] \ ,  \\ \nonumber V_{\pm}
& = & \frac{1}{2} \ (\tau_{4} \pm i \tau_{5}) \ , \ \ \ \ \
V_{3} =  \frac{1}{2} \ [V_{+}, V_{-}] \ .
\end{eqnarray}

\noindent The choice of derivations is dictated by which symmetries
we want unbroken at the
end. In electroweak theory electromagnetic $U_{em}(1)$ whose generator
is $\tau_{0} + \tau_{3}$ is unbroken. Among the above generators of
$M_{3}$ only the generators of the U-spin subalgebra commute with
$\tau_{0} + \tau_{3}$ so we define our
derivations as
\be\label{eqn2}
e_{a}(f) = m_{a} [\lambda_{a}, f], \ \ \ \ \ f \in M_{3}
\ee

\noindent where a runs through the indices (+, -, 3) and
\ben\label{eqn3}
\lambda_{\pm} & = & \frac{U_{\pm}}{\sqrt{2}} \ \ , \ \ \ \ \
\lambda_{3}  =  U_{3} \ ,
 \\ \nonumber
m_{\pm} & = & m   \ \ \ , \ \ \ \ \  \ m_{3}  =  \frac{m^{2}}{M} \ .
\een
\noindent Here m and M are two mass scales that have to be introduced
into the theory to keep the dimensions correct.
In defining the derivations we use the fact that all derivations of
$M_{n}$ are
inner and  hence they are in the form $e_{a} = ad(\lambda_{a})$. They
obey the
commutation relations
\be\label{eqn4}
[e_{a}, e_{b}] = \frac{m_{a} m_{b}}{m_{c}} \  C_{ab}^{c} \  e_{c}
\ee
\noindent where the structure constants $C_{ab}^{c}$ are
\be\label{eqn5}
C_{+-}^{3} = -C_{-+}^{3} = 1, \ \ \ \ \ C_{3+}^{-} = -C_{+3}^{-} = 1, \
\ \ \ \
C_{-3}^{+} = -C_{3-}^{+} = 1
\ee
\noindent and all others are zero.

We can now define the exterior derivative
exactly as in \cite{violette2}, but with the set of derivations in
Der$_{2}(M_{3}) \subseteq $Der($M_{3}$):
\be\label{eqn6}
df (e_{a}) = e_{a} (f).
\ee
\noindent This means in particular that
\be\label{eqn7}
d\lambda^{a} (e_{b}) = m_{b} \ [\lambda_{b}, \lambda^{a}],
\ee
\noindent where the indices are lowered and raised by the group metric
\be\label{eqn8}
g_{ab} = - Tr(\lambda_{a} \lambda_{b}).
\ee
We define the set of one forms $\Omega_{2}^{~1}(M_{3})$ to be the set of
all
elements of the form f dg or dg f with f and g in $\oz$ subject to the
relations $d(fg) = df g + f dg$. Here the subindex 2 refers to the fact
that we are
using the derivation algebra Der$_{2}(M_{3})$.
The set $d\lambda^{a}$ forms a system of generators of
$\Omega_{2}^{~1}(M_{3})$ as
a left or right module but it is not a convenient one since $\lambda^{a}
d\la^{b}
\neq d\la^{b} \la^{a}$. However there is another system of generators
completely
characterised by the equations

\begin{eqnarray}\label{eqn9}
  \theta_{\pm} (e_{\mp}) & = & 1 \ \ , \ \ \ \ \ \theta_{\pm} (e_{3}) =
  0 \ , \\ \nonumber
  \theta_{3} (e_{\mp}) & = & 0 \ \ , \ \ \ \ \ \theta_{3} (e_{3}) =
  1 \ .
\end{eqnarray}
They are related to $d\la^{a}$ by the equations
\be\label{eqn10}
d\la^{a} = m_{b} \ C^{~a}_{bc} \ \theta^{b} \ \la^{c}
\ee
\noindent and they satisfy the structure equations
\be\label{eqn11}
d\theta^{a} = C^{~a}_{bc} \ \frac{m_{b} m_{c}}{m_{a}} \ \theta^{b}
\wedge \theta^{c} \ \ .
\ee
\noindent The $\theta^{a}$'s commute with all elements of $M_{3}$.

Let us choose a basis $\theta^{\alpha}_{\beta} \ dx^{\beta}$ of
$\Omega^{1}(V)$ over V and suppose $e_{\alpha}$ be the pfaffian
derivations dual to $\theta^{\alpha}$. Set $i = (\alpha, a), 1 \  \leq
i  \ \leq \ 4 + 3 = 7 $ and introduce $\theta^{i} = (\theta^{\alpha},
\theta^{a})$ as generators of $\Omega^{1}(\oz)$ as a left or right
$\oz$-module and $e_{i} = (e_{\alpha}, e_{a})$ as a basis of
Der$_{2}(\oz)$ as a direct sum

\begin{equation}\label{eqn12}
  \Omega^{1}({\oz}) = \Omega^{1}_{h} \oplus \Omega^{1}_{v}
\end{equation}
where
\begin{equation}\label{eqn13}
  \Omega^{1}_{h} = \Omega^{1}(V) \otimes M_{n} \ \ \ \ \ , \ \ \ \
  \  \Omega^{1}_{v} = {\cal{C^{\infty}}}(V) \otimes \Omega^{1}(M_{n}) \
.
\end{equation}
Thus the exterior derivative df of an element f of ${\oz}$
can be written as the sum of its vertical and horizontal parts:
\begin{equation}\label{eqn14}
  df = d_{h}f + d_{v}f  \ \  .
\end{equation}

\noindent From the basis elements $\theta^{a}$ we can construct a 1-form
$\theta$ in $\Omega^{1}_{v}$, that is
\be\label{eqn15}
\theta = - m_{a} \ \la_{a}
\ee
\noindent which satisfies the zero-curvature condition
\be\label{eqn16}
d\theta + \theta^{2} = 0.
\ee

\section{Gauge Fields}
The gauge potential, which is an element of $\Omega^{1}(V)$ for a
trivial U(1)-bundle can be generalised to the noncommutative case as an
anti-Hermitian element of
$\Omega^{1}(\oz)$. Let $\omega$ be such an element of $\Omega^{1}(\oz)$.
We can
write it then as
\be\label{eqn17}
\omega = A + \theta + \Phi
\ee
\noindent where
\ben\label{eqn18}
A & = & - ig A_{\alpha} \theta^{\alpha}  \ \ \ \ \ \ \in
\Omega^{1}_{h}(\oz)\\ \nonumber
\Phi & = & g \phi_{a} \theta^{a} \ \ \ \ \ \ \ \ \ \  \in
\Omega^{1}_{v}(\oz)
\een
\noindent and $\theta$ as in ($\ref{eqn15}$). g is the coupling constant
of the theory. $\phi_{a}$ here are interpreted as Higgs fields.

The gauge transformations of the trivial U(1)-bundle over V are the
unitary elements of $C^{\infty}(V)$. In analogy, we will choose the
group of local gauge transformations as the group of unitary elements
${\cal{U}}$ of $\oz$, that is the group
of invertible elements  $u \in \oz$ satisfying $u u^{*} = 1$. Here * is
the *-product induced in $\oz$ and $\oz$ is considered as the set of
functions on V with values in $GL_{n}$.
An element of ${\cal{U}}$ defines a map of $\Omega^{1}({\oz})$ into
itself of the form

\begin{equation}\label{eqn19}
  \omega^{'} = g^{-1} \omega g + g^{-1} dg .
\end{equation}
We define
\begin{equation}\label{eqn20}
  \theta^{'} = g^{-1} \theta g + g^{-1} d_{v}g
\end{equation}
\begin{equation}\label{eqn21}
  A^{'} = g^{-1} A g + g^{-1} d_{h}g
\end{equation}
and so $\phi$ transforms under the adjoint action of
${\cal{U}}$:
\begin{equation}\label{eqn22}
  \phi^{'} = g^{-1} \phi g  \ \  .
\end{equation}

\noindent $\th$ is invariant under the gauge transformations and hence
$\omega^{'}$ is again of the form $(\ref{eqn17})$.
The curvature 2-form $\Omega$ and the field strength $F$ are defined as
usual
\begin{equation}\label{eqn23}
  \Omega = d\omega + \omega^{2}  \ \ \ \ \ \ \  F = d_{h}A + A^{2}
\end{equation}
with components
\begin{equation}\label{eqn24}
  \Omega = \frac{1}{2} \ \Omega_{ij} \theta^{i} \wedge \theta^{j}
  \ ,  \ \ \ \ F = \frac{1}{2} F_{\alpha \beta} \theta^{\alpha}
  \wedge \theta^{\beta} \ .
\end{equation}
We find
\begin{eqnarray}\label{eqn25}
\Omega_{\alpha \beta} & = &  F_{\alpha \beta}  \ \  ,
\\ \nonumber \Omega_{\alpha a} & = &  g {\cal{D}}_{\alpha} \phi_{a} =
 g ( \partial_{\alpha} \phi_{a} - ig \ [A_{\alpha},
\phi_{a}] )  \ \  ,
\\ \nonumber  & & \\ \nonumber \Omega_{ab} & = & g^{2} [\phi_{a},
\phi_{b}] - g \frac{m_{a} m_{b}}{m_{c}} \ C_{ab}^{~c} \ \phi_{c}.
\end{eqnarray}
As we shall see the term $\Omega_{ab}$ is responsible for the
Higgs potential.

Given the curvature 2-form, we can write down the usual gauge
invariant Yang-Mills Lagrangian density 4-form:
\be\label{eqn26}
{\cal{L}} = -\frac{1}{2 g^{2}} \ Tr(\Omega_{ij} \Omega^{ij})
\ee
\no In terms of the components of $\Omega$, ${\cal{L}}$ becomes
\be\label{eqn27}
{\cal{L}} = -\frac{1}{2 g^{2}} \ Tr(F_{\alpha \beta} F^{\alpha \beta}) -
Tr({\cal{D}}_{\alpha} \phi_{a} \ {\cal{D}}^{\alpha} \phi^{a}) + V(\phi)
\ee
\no where the Higgs potential $V(\phi)$ is given by
\be\label{eqn28}
V(\phi) = - \frac{1}{2 g^2} \ Tr(\Omega_{ab} \Omega^{ab})  .
\ee

\no From the form of $\Omega_{ab}$ in $(\ref{eqn25})$ we see that
$V(\phi)$ vanishes for values
\be\label{eqn29}
\phi_{a} = 0, \ \ \ \ \ \ \ \phi_{a} = \frac{m_{a}}{g} \ \la_{a} .
\ee
For the second vacuum configuration above, the second term on the right
hand side
of $(\ref{eqn27})$ becomes
\be\label{eqn30}
g^{2} \ Tr([A_{\alpha}, m_{a} \la_{a}] [A^{\alpha}, m_{a} \la^{a}]).
\ee
\no This expression is quadratic in the potential and hence it gives a
mass to the
vector bosons. This means we have a naturally built-in Higgs mechanism.

\section{The Higgs Masses}

In what follows we assume a Minkowskian space-time and work in Cartesian

coordinates. Therefore we take
$e_{\alpha} = \partial_{\alpha}$ and $\th^{\alpha} = dx^{\alpha}$. Hence
we have
\be\label{eqn31}
d_{h} = dx^{\alpha} \partial_{\alpha}
\ee
In this model there are three independent Higgs fields:
\be\label{eqn32}
\phi_{+} = \frac{H^{\dagger}}{\sqrt{2}}, \ \ \ \ \ \phi_{-} =
\frac{H}{\sqrt{2}}, \ \ \ \ \ \phi_{3} = \triangle + \frac{m^{2}}{2Mg} \
(2\tau_{0} - 1) .
\ee
where
\begin{eqnarray}\label{eqn34}
 H & = & H_{+} V_{+} + H_{0} U_{+} \ \  , \\ \nonumber
\triangle & = & \frac{1}{2} \ (\triangle_{0} \lambda_{0} +
\triangle_{a} \lambda_{a}) .
\end{eqnarray}

\no By using the metric components $(\ref{eqn8})$ we see that
\be\label{eqn33}
\phi^{+} = -2 \phi_{-},  \ \ \ \ \phi^{-} = -2  \phi_{+}, \ \ \ \
\phi^{3} = -2 \phi_{3} .
\ee

\no For the gauge potential we will write
\be\label{eqn35}
A  =  - ig A_{\mu} dx^{\mu} = - ig \frac{1}{2} \
(B_{\mu} \lambda_{0} + W_{\mu a} \lambda_{a}) dx^{\mu} ,
\ee
where $B$ and $W$'s are going to be identified as
the weak gauge bosons.

Using the field components above we can write the connection 1-form
directly
from $(\ref{eqn17})$:

\begin{eqnarray}\label{connection}
\omega & = & A + \frac{g}{\sqrt{2}} \ H \theta_{-} +
\frac{g}{\sqrt{2}} \  H^{\ast} \theta_{+} + g \triangle
\theta_{3}
\\ & & - \frac{m}{\sqrt{2}} \ U_{+} \theta_{-} -
\frac{m}{\sqrt{2}} \ U_{-} \theta_{+} + \frac{m^{2}}{4M} \
(\lambda_{0} + \lambda_{3}) \theta_{3}. \nonumber
\end{eqnarray}
The next step is to construct the curvature 2-form
\begin{eqnarray}\label{curvature}
  \Omega & = & \frac{1}{2} \ \Omega_{\mu \nu} dx^{\mu} dx^{\nu} +
  \Omega_{\mu +} dx^{\mu} \theta_{-} + \Omega_{\mu -} dx^{\mu}
  \theta_{+} + \Omega_{\mu 3} dx^{\mu} \theta_{3} \\ \nonumber
  & & + \Omega_{+ -} \theta_{-} \theta_{+} + \Omega_{+ 3}
  \theta_{-} \theta_{3} + \Omega_{3 -} \theta_{3} \theta_{+}.
\end{eqnarray}
From $(\ref{eqn25})$ we can see that
\begin{eqnarray}\label{eqn36}
\Omega_{\mu \nu} & = &  F_{\mu \nu} , \\ \nonumber \Omega_{\mu +} &
= & \frac{g}{\sqrt{2}} \ {\cal{D}}_{\mu} H \ \ , \ \ \ \ \ \
\Omega_{\mu - } = \Omega^{\ast}_{\mu +} \ \ , \\ \nonumber \Omega_{\mu
3} & = & g {\cal{D}}_{\mu} \triangle
\end{eqnarray}
where
\begin{equation}\label{eqn37}
  {\cal{D}}_{\mu} = \partial_{\mu} -i g [A_{\mu}, \quad ]
\end{equation}
and the remaining three terms are
\begin{eqnarray}\label{eqn38}
\Omega_{+ -} & = & \frac{g^{2}}{2} \ [H, H^{\ast}] - g M
\triangle - m^{2} \lambda_{0} + \frac{m^{2}}{2} , \\ \nonumber
\Omega_{+ 3} & = & - \frac{g^{2}}{\sqrt{2}} \ \triangle H \ \
\ \ , \ \ \ \Omega_{3 -} = \Omega_{+ 3}^{\ast} .
\end{eqnarray}
These can also be found directly from (3.9) and the definitions
(4.2) and (4.4).
We write down the Lagrangian as before and obtain
\be\label{lagrangian}
{\cal{L}} = -\frac{1}{2 g^{2}} \ Tr(F_{\alpha \beta} F^{\alpha \beta}) +
2Tr({\cal{D}}_{\alpha} H \ {\cal{D}}^{\alpha} H^{\dagger}) +
2Tr({\cal{D}}_{\alpha} \Delta \ {\cal{D}}^{\alpha} \Delta^{\dagger}) +
V(H, \triangle),
\ee
where the Higgs potential is
\begin{eqnarray}\label{eqn39}
  \frac{1}{8 g^{2}} \ V(H, \triangle) & = & \frac{1}{8} \
\left[H^{\dagger}H
  - \frac{m^{2}}{g^{2}}\right]^{2} \\ \nonumber
  & & + \frac{1}{4} \ \left[\frac{1}{2} \ H^{\dagger}H - \frac{M}{g} \
  \triangle_{0} - \frac{m^{2}}{g^{2}}\right]^{2} \\ \nonumber
  & & + \frac{1}{4} \ \left[\frac{1}{2} \ H^{\dagger} \sigma_{a} H -
  \frac{M}{g} \ \triangle_{a}\right]^{2} \\ \nonumber
  & & + \frac{1}{8} \ H^{\dagger} (\triangle_{0} + \triangle_{a}
  \sigma_{a})^{2} H .
\end{eqnarray}
\no Above H is written as a two-component
column vector with complex entries $H_{+}$ and $H_{0}$ and
$\sigma_{a}$ are the Pauli spin matrices.
The vacuum configuration can be determined either directly from the
minimum
of the
above potential which is a sum of squares, or from $(\ref{eqn29})$, to
be

\begin{equation}\label{eqn40}
  H_{0} = \frac{m}{g}  \ \ \ \ H_{+} = 0  \ \ \ \ \triangle_{0} =
  \triangle_{3} = - \frac{m^{2}}{2Mg}  \ \ \ \ \triangle_{1,2} = 0
\end{equation}
where only the electromagnetism survives symmetry breaking.

In this model we have considered our structure group
$SU(2)_{I} \times U(1)_{Y}$ as a
subgroup of $U(3)$ and hence their coupling constants $g$ and $g'$
merge to the same value .
As a consequence, the Weinberg angle is obtained from the
standard relation $$\sin^{2}\theta_{w} = \frac{g^{2}}{g^{2} +
g'^{2}} = \frac{1}{2}.$$
The mass spectrum of the model can be found easily. The masses of
the W and Z bosons are found from $(\ref{eqn30})$ to be
 $$M_{W} = m \, \sqrt{1 + \frac{m^{2}}{2M^{2}}} \quad \quad
M_{Z} = \sqrt{2} m$$
To find the mass spectrum of the Higgs sector on the other hand,
we first write down the linearized
 field equations \cite{catal}:
\begin{eqnarray}\label{field}
d\star dH_{1} & + & 2 m^{2} \ \left(1 + \frac{m^{2}}{2M^{2}}
\right)H_{1}
- 2Mm \  \left(1 + \frac{m^{2}}{2M^{2}}\right)
\triangle_{1}  =  0
\\ \nonumber
d\star dH_{2} & + & 2 m^{2} \
\left(1 + \frac{m^{2}}{2M^{2}} \right)H_{2} + 2Mm \
\left(1 + \frac{m^{2}}{2M^{2}}\right)
\triangle_{2}  =  0
\\ \nonumber
 d\star dH_{3} & + & 8 m^{2} H_{3} - 2Mm \
(\triangle_{0} - \triangle_{3})  =  0 \\ \nonumber
d\star dH_{4}   &   &  = 0 \\ \nonumber
 d\star
d\triangle_{0} & - & 2Mm \ H_{3} + 2 M^{2} \  \left(1 +
\frac{m^{2}}{2M^{2}}\right) \triangle_{0} -
m^{2} \triangle_{3}  = 0 \\ \nonumber
 d\star d\triangle_{3}  & + & 2Mm \
H_{3} + 2 M^{2} \ \left(1 + \frac{m^{2}}{2M^{2}}\right)
\triangle_{3} - m^{2} \ \triangle_{0}  =  0 \\ \nonumber
d\star d\triangle_{1} & + & 2 M^{2} \ \left(1 +
\frac{m^{2}}{2M^{2}}\right) \triangle_{1} - 2Mm \ \left(1 +
\frac{m^{2}}{2M^{2}}\right) H_{1}  =  0 \\ \nonumber
 d\star
d\triangle_{2} & + & 2 M^{2} \ \left(1 + \frac{m^{2}}{2M^{2}}\right)
\triangle_{2} + 2Mm \ \left(1 +
\frac{m^{2}}{2M^{2}}\right) H_{2} =  0
\end{eqnarray}
\no where
$$H_{1} = H_{+} + H_{+}^{*}, \ \ \ \ \ \ \ H_{2} = (H_{+} - H_{+}^{*})/i
,$$
$$H_{3} = H_{0} + H_{0}^{*}, \ \ \ \ \ \ \ H_{4} = (H_{0} - H_{0}^{*})/i
.$$

\no
The diagonalization of the mass matrix that is read from linearized
Higgs field equations yields the mass eigenvalues
$$0, \, 0, \, 0, \, 2 M^2, \quad
( 3m^2 + \frac{m^4}{M^2} + 2M^2 ), \quad
( 3m^2 + \frac{m^4}{M^2} + 2M^2 ),$$

$$( 5m^2+ M^2 + \sqrt{9m^{4}+ 2 m^2 M^{2} + M^4} ), \quad \quad
( 5m^2+ M^2 - \sqrt{9m^{4}+ 2 m^2 M^{2} + M^4} ).$$
corresponding to the Higgs scalars
$$H_{+} \pm H_{+}^{*}, \ \ \ \ \ H_{0} \pm H_{0}^{*},\ \ \ \\
\triangle_{1} \pm i \triangle_{2}, \ \ \ \ \ \triangle_{0} \pm
\triangle_{3}.$$
The value of the Weinberg angle and the above masses imply that
the $\rho$ parameter  $$\rho = \frac{M_{W}^{2}}{M_{Z}^{2}
\cos^{2}\theta_{W}} = 1 + \frac{m^{2}}{2M^{2}}.$$ Experimentally
$\rho$ is very close to one so we must have $M \gg m$.
 Thus at the mass scale $M$ we obtain
three zero mass eigenvalues that refer to Goldstone modes
which would be absorbed by
weak intermediate bosons to become massive,
one {\sl light} Higgs boson with mass
$\sqrt{2}m$,
and all the remaining scalar masses converge to $\sqrt{2}M$ as we take
M $\gg$ m.

It is now possible to predict  the values of these masses
at the electroweak scale $E_Z \sim m$
by considering the
renormalization flow of the coupling constants $g$, $g'$  and  the Higgs

self-coupling constant $\lambda$ down from the scale $M$ to the scale
$m$ and also
using the fact that $\lambda = \frac{g^2}{4}$ from the Higgs potential
$(\ref{eqn39})$ $\cite{feza2}$.
The relevant renormalisation group equations  are given by
\cite{schrempp}
\begin{equation}\label{eqn210}
  16 \pi^{2} \frac{dg}{dt} = - \frac{19}{6} g^{3}  ,
\end{equation}
\begin{equation}\label{eqn220}
16 \pi^{2} \frac{dg'}{dt} =  \frac{41}{6} g'^{3}  ,
\end{equation}
\begin{equation}\label{eqn230}
  16 \pi^{2} \frac{d\lambda}{dt} = 24 \lambda^{2} - 3 \lambda (3
  g^{2} + g'^{2}) + \frac{3}{8}  \ [2 g^{4} + (g^{2} + g'^{2})^{2}] .
\end{equation}
We  solve ($\ref{eqn210}$) and
($\ref{eqn220}$)  and  set  $g = g'$ and $\lambda = \frac{1}{4} \
g^{2}$ at the scale $M$. This implies
\begin{equation}\label{eqn240}
  \frac{1}{g^{2}(\mu)} - \frac{1}{g'^{2}(\mu)} = \frac{60}{48
  \pi^{2}} \ \ln \frac{\mu}{M}
\end{equation}
at an arbitrary mass scale  $\mu$. We fix
 $g$ and $g'$ at the scale $\mu = E_{Z}= 91 GeV$
by their measured values
$g(E_{Z}) = 0.4234$ and $g'(E_{Z}) = 0.1278$.
This choice   drives the Weinberg
angle to its experimental value 0.23 at the scale $\mu = E_{Z}$.
We also  find that we should  have $M \sim
5 \times 10^{20} GeV$ to start with.
Inserting what we found  back into $(\ref{eqn210})$ and
($\ref{eqn220}$) we obtain
\begin{equation}\label{eqn250}
  g^{2}(M) = g'^{2}(M) = 4 \lambda (M) = 0.49 .
\end{equation}
The remaining equation
($\ref{eqn230}$) can  be solved numerically by feeding in
the solutions of  ($\ref{eqn210}$) and ($\ref{eqn220}$)
yielding  the result  $\lambda(E_{Z}) = 0.14$.
From the standard model
\begin{equation}\label{eqn260}
  \frac{m_{H}^{2}(\mu)}{m_{Z}^{2}(\mu)} = \frac{8
  \lambda(\mu)}{g^{2}(\mu) + g'^{2}(\mu)}
\end{equation}
which is already satisfied at scale $M$.
This relation gives  the mass of the Higgs particle
at the electroweak scale as  $m_{H}(E_{z}) = 130GeV$.
However, the actual determination of the  physical mass
should take into consideration  radiative corrections.
But it is well known that \cite{sirlin}
\begin{equation}\label{eqn270}
  m_{H}(\mu) = m_{H}^{pole} (1 + \delta(\mu))
\end{equation}
where $\delta(\mu)$ referring to the radiative corrections
are very small at the scale $\mu = E_{Z}$.
Therefore we may  conclude $m_{H}^{pole} \sim m_{H}(E_{Z}) \sim 130GeV$.

\section{Concluding Comments}

The non-commutative extension of the electroweak model proposed by
Balakrishna, G\"{u}rsey and Wali \cite{feza2}  where the space-time is
extended by the Pauli matrices themselves
is both intiutive and comparatively simple to study.
It predicts  a {\sl light} Higgs boson with a mass around $130 GeV$
together with four very heavy Higgs bosons.

The model may be generalized in several directions.
In fact a supersymmetric generalization \cite{feza1}
as well as a grand unification scheme \cite{feza3}
had already been discussed. We think it wouldn't be unreasonable to
contemplate
an effective field theory approach  to the non-commutative electroweak
models.
In a first attempt, we consider to the lowest possible order, the
following
cubic term in the Higgs potential:

\be
\frac{\alpha}{ 3! g^{3}} \ Tr(\Omega_{ab}  \ \Omega^{bc} \  \Omega_{cd}
\ g^{ad})
\ee
\no which contributes as
\ben\label{eqn100}
& & \frac{\alpha g^{3}}{4} \  ([\sum_{i=0}^{3} \ \triangle_{i}
(H^{\dagger} \sigma_{i} H) ]^{2} + H^{\dagger} H [H^{\dagger}
(\sum_{i=0}^{3} \ \triangle_{i} \sigma_{i} )^{2} H]) \\ \nonumber
& + & \frac{\alpha g^{2}}{4} \ M \ (H^{\dagger} (\sum_{i=0}^{3} \
\triangle_{i} \sigma_{i} )^{3} H) - \frac{\alpha g}{2} \ m^{2}
H^{\dagger} (\sum_{i=0}^{3} \ \triangle_{i} \sigma_{i})^{2} H.
 \een
It can be checked that the vacuum configuration $(\ref{eqn40})$ makes
the above expression vanish.
This doesn't necessarily mean
that with the inclusion of effective terms, the complete Higgs potential

cannot acquire a distinct set of vacuum expectation values.
The possibility remains open at present.

\section{Acknowledgement}
We are grateful to Professor K. C. Wali for discussions.
We thank T\"{U}B\.{I}TAK (Scientific and Technical Research Council of
Turkey)
for support through the BAYG-BDP program.

\bigskip



\end{document}